\documentclass[]{spie}  
\usepackage[]{graphicx}

\title{Lyot-based Low Order Wavefront Sensor: Implementation on the Subaru Coronagraphic Extreme Adaptive Optics System and its Laboratory Performance} 

\author{Garima Singh\supit{a,b}, Olivier Guyon\supit{a,d}, Pierre Baudoz\supit{b}, Nemanja Jovanovic\supit{a}, Frantz Martinache\supit{c}, Tomoyuki Kudo\supit{a}, Eugene Serabyn\supit{d}, Jonas G Kuhn\supit{d}
\skiplinehalf
\supit{a}National Astronomical Observatory of Japan, Subaru Telescope, 650 North A'Ohoku Pl, Hilo-96720, Hawaii, USA; \\
\supit{b}Laboratoire d'Etudes Spatiales et d'Instrumentation en Astrophysique, Observatoire de Paris-Meudon, 5 Place Jules Janssen, F-92195 Meudon, France;\\
\supit{c}Laboratoire Lagrange, UMR7293, Universit\'e de Nice Sophia-Antipolis, CNRS, Observatoire de la C\^ote d'Azur, Bd. de l'Observatoire, 06304 Nice, France;\\
\supit{d}Jet Propulsion Laboratory, California Institute of Technology, 4800 Oak Grove Drive, Pasadena, CA 91109-8099, USA}

\authorinfo{Further author information: Send correspondence to Garima Singh, E-mail: singh@naoj.org, Telephone: +1 808 934 5966}
\begin{document} 
\maketitle 

\begin{abstract}
High throughput, low inner working angle (IWA) phase masks coronagraphs are essential to directly image and characterize (via spectroscopy) earth-like planets. However, the performance of low-IWA coronagraphs is limited by residual pointing errors and other low-order modes. The extent to which wavefront aberrations upstream of the coronagraph are corrected and calibrated drives coronagraphic performance.
Addressing this issue is essential for preventing coronagraphic leaks, thus we have developed a Lyot-based low order wave front sensor (LLOWFS) to control the wavefront aberrations in a coronagraph. The LLOWFS monitors the starlight rejected by the coronagraphic mask using a reflective Lyot stop in the downstream pupil plane. The early implementation of LLOWFS at LESIA, Observatoire de Paris demonstrated an open loop measurement accuracy of 0.01 $\lambda$/D for tip-tilt at 638 nm when used in conjunction with a four quadrant phase mask (FQPM) in the laboratory. To further demonstrate our concept, we have installed the reflective Lyot stops on the Subaru Coronagraphic Extreme AO (SCExAO) system at the Subaru Telescope and modified the system to support small IWA phase mask coronagraphs ($< 1\lambda$/D) on-sky such as FQPM, eight octant phase mask, vector vortex coronagraph and the phase induced amplitude apodization complex phase mask coronagraph with a goal of obtaining milli arc-second pointing accuracy. Laboratory results have shown the measurement of tip, tilt, focus, oblique and right astigmatism at 1.55 $\mu$m for the vector vortex coronagraph. Our initial on-sky result demonstrate the closed loop accuracy of $<$ 7 x 10$^{-3}$ $\lambda/$D at 1.6 $\mu$m for tip, tilt and focus aberrations with the vector vortex coronagraph. 
\end{abstract}



\section{INTRODUCTION}
\label{sec:intro}  

Direct detection of exoplanets is one of the most challenging field in Astronomy today. To image an Earth around the sun from a distance of 10 pc requires a contrast of 10$^{-10}$ in the visible and 10$^{-7}$ in the thermal IR. Such high contrasts from ground-based telescopes are limited by many factors such as: distortion of the wavefront by the atmospheric turbulence which reduces the peak intensity of the star and creates a halo of speckles around the point spread function (PSF), the optical imperfections of the imaging instrument, the local turbulence induced due to variations in temperature and the residual quasi-static speckles. 

Eight meter class telescopes equipped with high performance adaptive optics (AO) systems are capable of suppressing the speckle noise by providing typical wavefront residuals of $<$ 200 nm rms at H band. But to achieve a raw contrast of $\approx$ 10$^{-4}$ and a detection contrast of at least  $\approx$ 10$^{-7}$ at small angular resolution ($\approx$ 1 $\lambda$/D) requires high performance low inner working angle (IWA) phase mask coronagraphs (PMCs) and an extreme adaptive optic system (ExAO) to provide rms wavefront errors of $<$ 50 nm. But how efficiently an ExAO system corrects and calibrates the wavefront aberrations occurring before the coronagraph, derives the performance of the direct imaging instruments. The lack of fine pointing control puts light in the 1 to 2 $\lambda$/D region of the focal plane, making it difficult to differentiate planet signals from the starlight leakage which has a direct impact on the sensitivity of the low IWA phase mask coronagraph.\cite{1}$^{,}$\cite{2}$^{,}$\cite{3}$^{,}$\cite{4}$^{,}$\cite{5}

\begin{figure}[h]
   \begin{center}
   \begin{tabular}{c}
   \includegraphics[height=6cm]{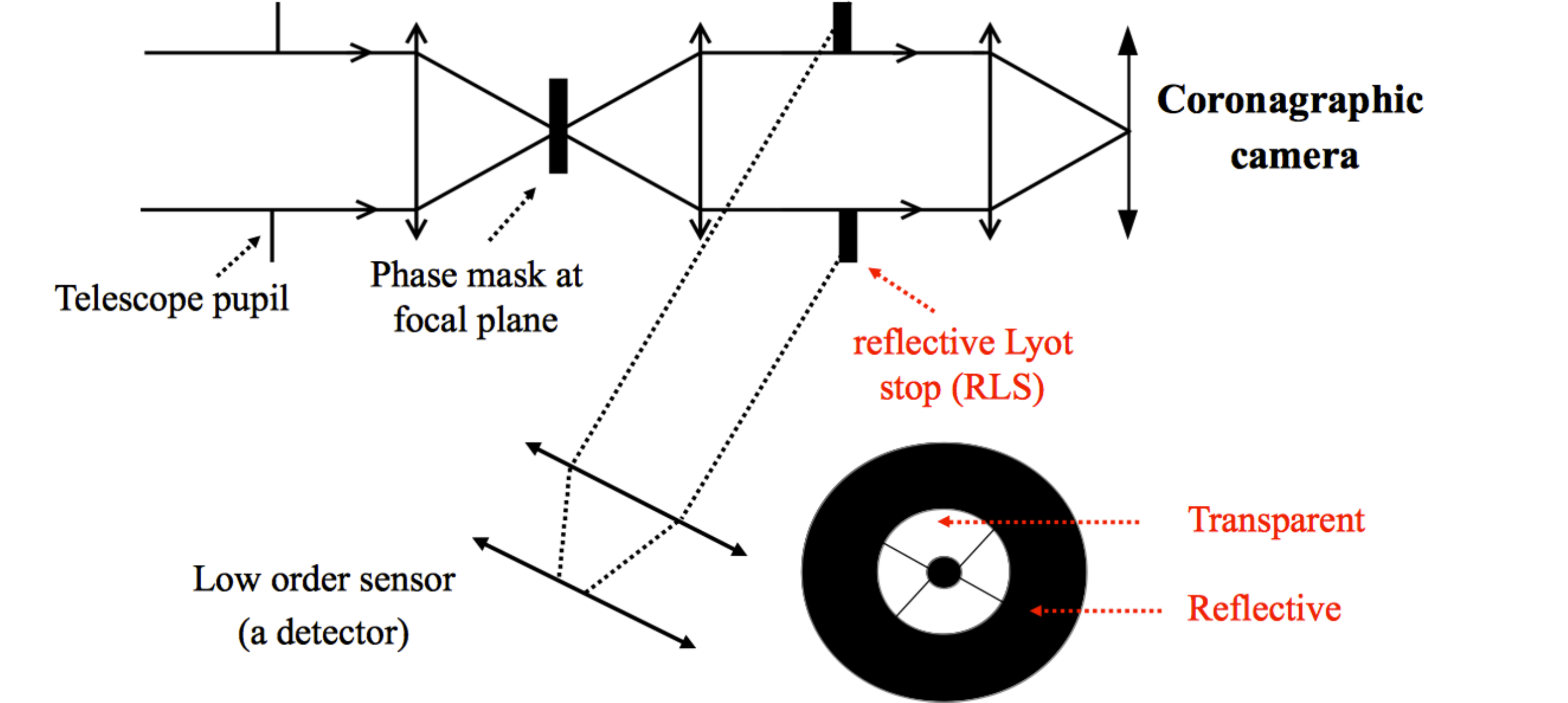}
   \end{tabular}
   \end{center}
   \caption[example] 
   { \label{fig:fig1} 
Basic schematic diagram of the Lyot-based low order wavefront sensor (LLOWFS). The phase mask at the focal plane diffracts starlight which is then reflected by the reflective Lyot stop (RLS) towards a reimaged focal plane. The low order sensor measures the low order aberrations. Note that the low order sensor is at a defocused position.\cite{6}$^{,}$\cite{7} The defocus is introduced so that focus aberration can be detected accurately.}
\end{figure}

To deal with the pointing errors occurring in the phase mask coronagraphs, we have developed a Lyot-based low order wavefront sensor\cite{6} (LLOWFS) illustrated in Figure~\ref{fig:fig1}. The AO-corrected beam enters the telescope pupil and encounters a phase mask at the focal plane. The phase mask diffracts the starlight outside of the geometrical pupil, which in the case of a conventional coronagraph, is absorbed by the Lyot stop in a reimaged pupil plane. With the LLOWFS approach, this unused diffracted starlight is reflected via a reflective Lyot stop (RLS) towards a reimaged focal plane. The reflected light is then sensed by the low order sensor (a detector) which measures the low order wavefront aberrations.  

LLOWFS is the next generation of coronagraphic low order wave front sensor\cite{7} (CLOWFS) which used a partially reflective focal plane mask to control pointing errors for occulting coronagraphs only. Although CLOWFS provided a high level of tip-tilt control, it was not compatible with phase mask coronagraphs as phase masks are not reflective. 

In our previous work\cite{6} we outlined the LLOWFS's principle, simulations and laboratory experiment in conjuction with a four quadrant phase mask\cite{8}. The paper demonstrates tip/tilt measurement with an accuracy of 0.01 $\lambda$/D in a coronagraphic testbed at LESIA, Observatoire de Paris. LLOWFS is a linear wavefront reconstructor which relies on the assumption that if post AO corrected wavefront residual are $<$ 1 radian rms then the intensity variations in the reflected light are linearly proportional to the low order aberration occurring before the focal plane phase mask.

In this paper, we demonstrate the implementation and laboratory experiments of LLOWFS on Subaru Telescope's high contrast imaging testbed known as SCExAO\cite{13}, the Subaru Coronagraphic Extreme Adaptive Optics system. In this body of work, we will present the LLOWFS capability of measuring the low order aberrations such as tip, tilt, focus, oblique astigmatism \& right astigmatism with the vector vortex coronagraph (VVC\cite{9}). In section 2, the SCExAO testbed is explained briefly. The integration of LLOWFS on SCExAO is illustrated in section 3. The procedure of measuring the pointing errors is defined in section 3.1. We present the laboratory and early on-sky closed loop results with the VVC in section 4. The LLOWFS performance on SCExAO is summarized in section 5. 
\section{SCExAO Overview} 
\label{sec:scexao}

The Subaru Coronagraphic Extreme AO (SCExAO\cite{12}) system is a highly flexible, high performance coronagraphic imaging system that uses intermediate remapping optics (phase induced amplitude apodization\cite{10} (PIAA) and Inverse PIAA) and is capable of detecting exoplanets as close as 1~$\lambda$/D. 

\begin{figure}[h]
   \begin{center}
   \begin{tabular}{c}
   \includegraphics[height=7cm]{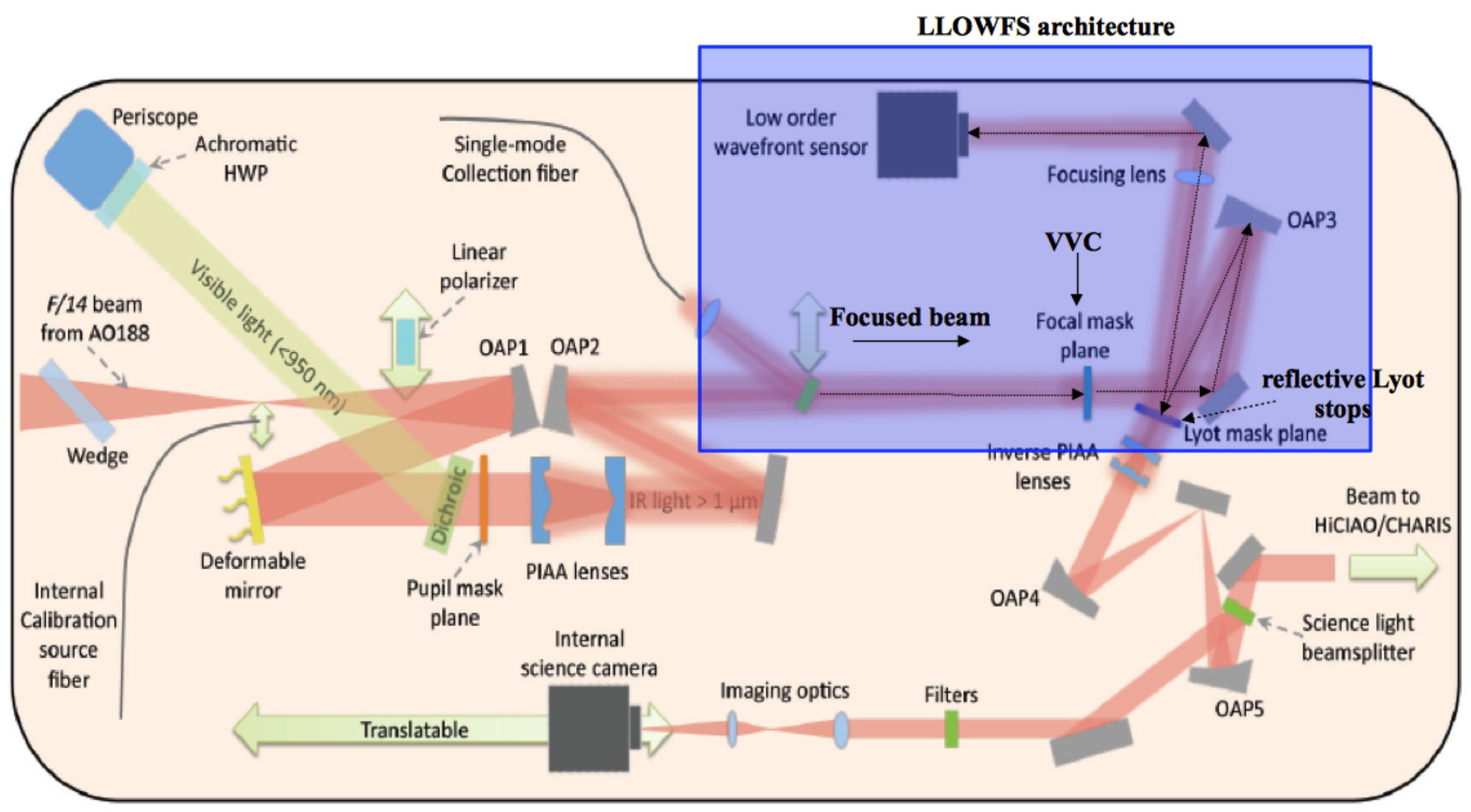}
   \end{tabular}
   \end{center}
   \caption[example] 
   { \label{fig:fig2} The schematic design of the infrared lower bench of SCExAO\cite{12}. The highlighted area shows the configuration of LLOWFS. The f/28 focused beam encounters the VVC which is installed in one of the slots of the focal plane mask wheel. The light rejected by the VVC then encounters the Lyot mask plane wheel where the reflective Lyot stops are situated.  The reflected light is directed towards the low order sensor via a focusing lens. The sensor then measures the low order aberrations.
}
\end{figure}

SCExAO is fed by the 188 element AO facility (AO188) of Subaru Telescope which typically provides the residual wavefront of $\approx$ 200 nm rms at H band. SCExAO has two levels: Visible upper bench which uses the light within wavelength 600~nm - 930~nm; Infrared lower bench (Figure~\ref{fig:fig2}) which uses the light between 950~nm - 2500~nm. SCExAO consists of numerous modules including a Pyramid wavefront sensor to measure high order aberrations in the visible at 750~nm; Lyot-based low order wavefront sensor to measure pointing errors and other low order aberrations at 1600~nm; a speckle nulling algorithm to suppress the speckle noise and a 2000 actuator MEMS Deformable Mirror (DM) to correct for high and low order aberrations. The output of SCExAO is fed to the HAWAII 2RG detector in HICIAO\cite{11} (High contrast instrument for the Subaru Next generation AO) for deep imaging of the post coronagraphic images.

The SCExAO bench has recently been equipped with phase mask coronagraphs such as VVC, FQPM, eight octant phase mask\cite{13} (8OPM) and the phase induced amplitude apodization complex phase mask coronagraph (PIAACMC\cite{14}) in order to reach an IWA $< 1~\lambda$/D. The corresponding reflective Lyot stops have also been installed on SCExAO. Jovanovic et al. (2014)\cite{12} offers a detailed explanation of the SCExAO instrument.
\section{LLOWFS Implementation on SCExAO} 
\label{sec:lowfs}

As shown in Figure~\ref{fig:fig1}, LLOWFS essentially requires a focal plane phase mask, a reflective Lyot stop, focusing optics and a near infrared camera.  The various components of LLOWFS implemented in the laboratory on SCExAO are:

\begin{itemize}
\item{Laser source of $\lambda$~=~ 1550~$nm$ with 15~$nm$ bandwidth}
\item{Subaru pupil mask}
\item{VVC (optimized at $\lambda$~=~1650~$nm$), mounted in a focal plane mask wheel which has degrees of freedom in the x, y and z directions. The VVC installed on SCExAO is Liquid Crystal Polymer (LCP)-based, sandwiched in A/R coated glass and is manufactured by JDSU. There is a 25 $\mu$m size metallic spot at the center of the VVC covering the central disorientation (defect) region of the vortex.}
\item{RLS (Lyot outer pupil is 8.6 mm diameter with a reflective annulus of 25.4~mm around it) mounted in a Lyot stop wheel (degrees of freedom: x and y directions), placed at an angle of 8$^{o}$ to reflect the light towards the low order sensor. 

The RLS is a fused silica disk of 1.5 mm thickness as shown in Figure~\ref{fig:fig3}(a). The substrate flatness is better than 5~$\mu$m. The black region is reflective chrome with 60~\% reflectivity in near IR (1200 nm) while the white region is transparent}
\item{Low-order sensor (320 x 256 pixels, InGaAs, pixel size: 30~$\mu$m, read out noise (e$^{-}$): 150, frame rate: 170~Hz). We introduced defocus in the sensor position and we estimated the value to be 5 radians rms. } 
\end{itemize}

Figure~\ref{fig:fig2} represents the optical design of SCExAO's infrared bench. A 1500 nm diverging laser beam from the internal calibration source is incident on an off-axis parabolic mirror (OAP1) creating an 18 mm collimated beam. The reflected beam is incident on the 2000 actuator MEMS DM which is conjugated to the telescope pupil. The beam is reflected towards a fixed Subaru pupil mask (with a central obscuration and spider arms) which is positioned as close to the pupil plane as possible. The beam then encounters a dichroic which lets IR~(~$>$~950~nm) to pass through the rest of the system.

There are two mask wheels immediately after the pupil mask which are also placed as close to the DM pupil as possible. These include the aspheric optics to apodize the beam for the PIAA coronagraph. The mask wheel can be moved in and out of the transmitted beam as per the coronagraph requirements. The flat mirror steers the beam onto OAP2 which focuses the beam on the focal plane mask wheel. This contains numerous phase mask coronagraphs as mentioned earlier. The VVC is situated in one of the wheel's slots. The coronagraphic beam after getting diffracted from a phase mask then encounters OAP3 which recollimates the beam to a 9 mm diameter. The collimated coronagraphic beam is incident on the Lyot stop wheel which is in a plane conjugated to the pupil. The reflective Lyot stop transmits the off axis source towards a science camera and reflects the on-axis diffracted starlight towards a focusing lens which reimages the reflective beam onto a low order sensor. We will concentrate only on LLOWFS architecture in this paper.
\subsection{Procedure of measuring the aberrations} 
\label{sec:results}

\begin{figure}[h]
   \begin{center}
   \begin{tabular}{c}
   \includegraphics[height=8cm]{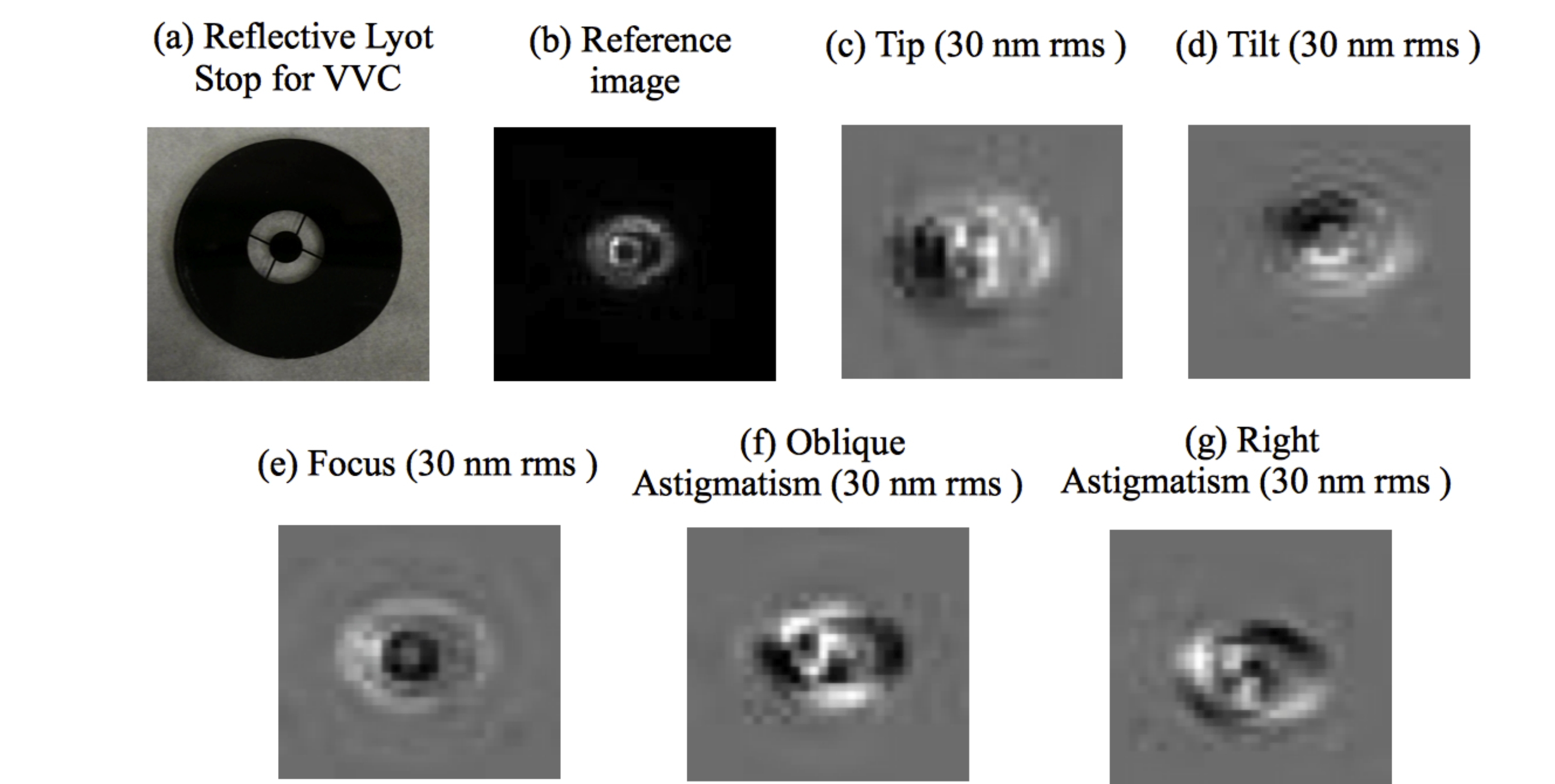}
   \end{tabular}
   \end{center}
   \caption[example] 
   { \label{fig:fig3} Images acquired on SCExAO bench in the laboratory. (a) The reflective Lyot stop for VVC. (b) Reference image with no aberration applied. The images from $( c)$ to (g) are the reference subtracted calibration frames. These differential images are called as the response of the sensor to the low order modes which are obtained by sending 30 nm rms phasemap (60 nm rms on the wavefront) to the DM for $( c)$ Tip mode. (d) Tilt mode. (e) Focus mode. (f) Oblique astigmatism mode. (g) Right astigmatism mode. We use these calibration frames to measure the low order aberrations in the aberrated wavefront at the entrance pupil. Note: All the calibration frames are of same brightness scale. 
}
\end{figure}

LLOWFS is a differential tip tilt sensor\cite{6}, therefore it requires calibration prior to measuring the aberrations in the wavefront. We acquire the response matrix for tip, tilt, focus, oblique and right astigmatisms by applying known amounts of each of these aberrations to our system independently. Figure~\ref{fig:fig3} (b) shows the reference image ($I_{0}$) acquired with no aberration. Figure~\ref{fig:fig3} (c-f) show the calibration frame acquired by applying phasemap with 60~nm rms wavefront error to the DM for each mode as mentioned.

Considering only tip/tilt, the LLOWFS is summarized in equation:  
\begin{equation}
\noindent
I _{R(\alpha_{x},\alpha_{y})} - I_{0} = \alpha_{x} S_{x} + \alpha_{y} S_{y}   
\label{eq:1}
\end{equation}

where $S_{x}$ and $S_{y}$ represent the sensor's response to tip and tilt respectively. For any instant image $I_{R}$ reflected by the RLS, one can identify the unknown tip-tilt $(\alpha_x,\alpha_y)$, by using a least squares algorithm. Equation \ref{eq:1} can be modified by adding more modes and its possible to measure other low order modes more than just tip-tilt.

In the SCExAO system, we applied tip aberration between $\pm$~85~nm rms (170~nm rms on the wavefront) with a step of 1 nm rms. This was done by sending phasemaps to the DM. For each phasemap applied to the DM, the reflected image $I_{R}$ is recorded. The tip aberration for each $I_{R}$ is then estimated through equation \ref{eq:1}. 

The same procedure is repeated for other modes such as tilt, focus, oblique and right astigmatisms. We discuss the response of the sensor to the aberrations applied in the next section. 
\section{LLOWFS Results} 
\label{sec:results}
\subsection{Laboratory Results} 
\label{sec:lab}
In this paper the LLOWFS's capability to measure the tip-tilt and other low order modes is presented. We intend to discuss two properties of the LLOWFS: linearity in the sensor measurement and the cross coupling between the modes. The linearity is defined as the maximum deviation of the measured aberration from the line of best fit. The fit is obtained through linear regression.  The accuracy of the linear fit is calculated as the ratio of the maximum output deviation divided by the full scale output, specified as a non linearity percentage. 
 
\begin{figure}
   \begin{center}
   \begin{tabular}{c}
   \includegraphics[height=8.4cm]{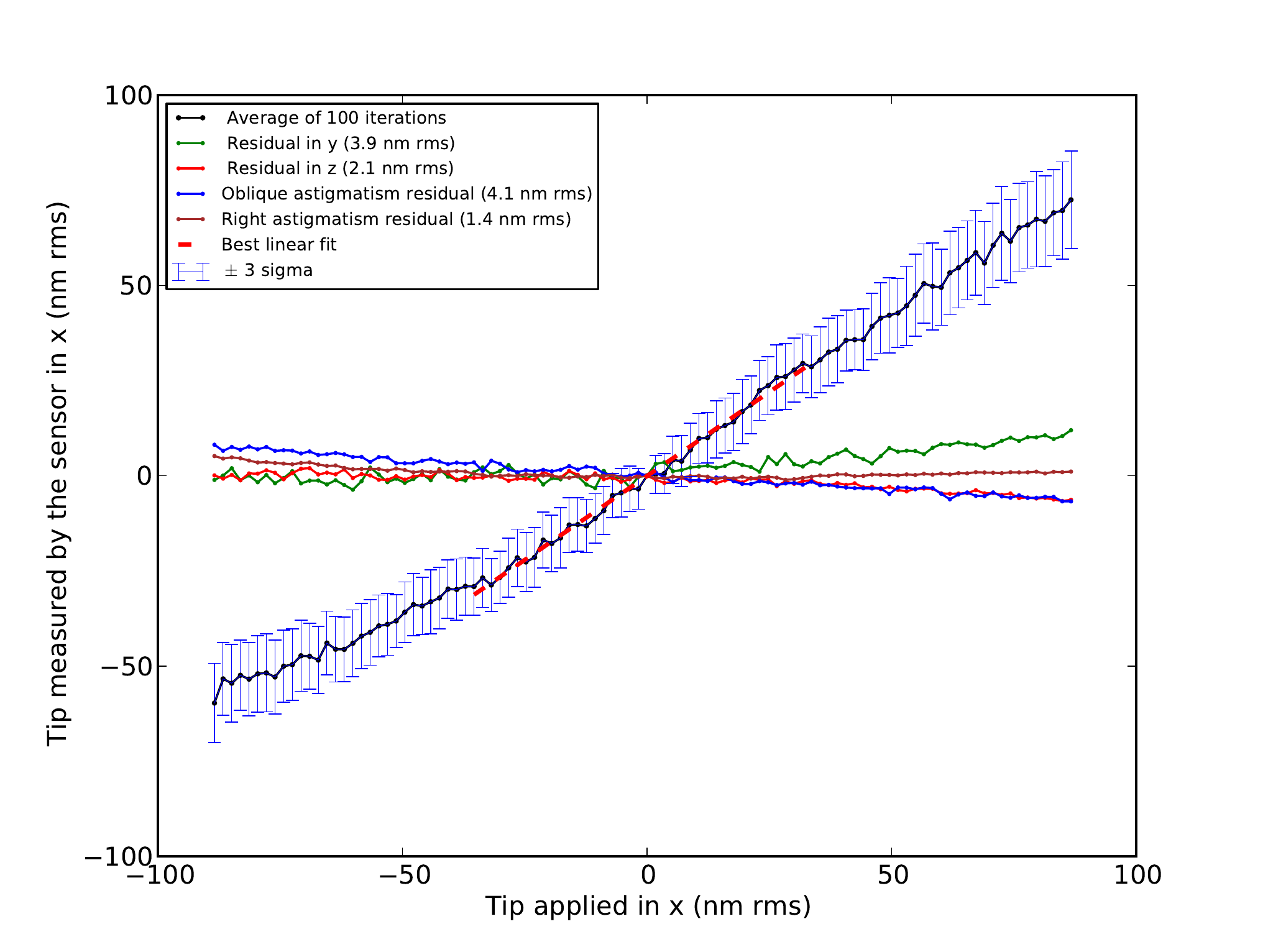}
   \end{tabular}
   \end{center}
   \caption[example] 
   { \label{fig:fig4} Linear response of the sensor to the Tip/Tilt aberration: X-axis plots the tip within $\pm$~85~nm rms applied to the DM ($\pm$ 170 nm rms on the wavefront ). Y-axis plots the tip measured by the sensor. The response of the sensor is 4.7\% non-linear over the range of $\pm$ 35~nm rms. The red dash line shows the best linear regression over the linear range of $\pm$ 35~nm rms. The residuals of the other low order aberrations average around zero indicating that there is no cross coupling. The experiment is repeated 100 times and the dispersion of the sensor measurements is thus shown as a $\pm~3 \sigma$ error bar.}
\end{figure}

\begin{figure}
   \begin{center}
   \begin{tabular}{c}
   \includegraphics[height=8.4cm]{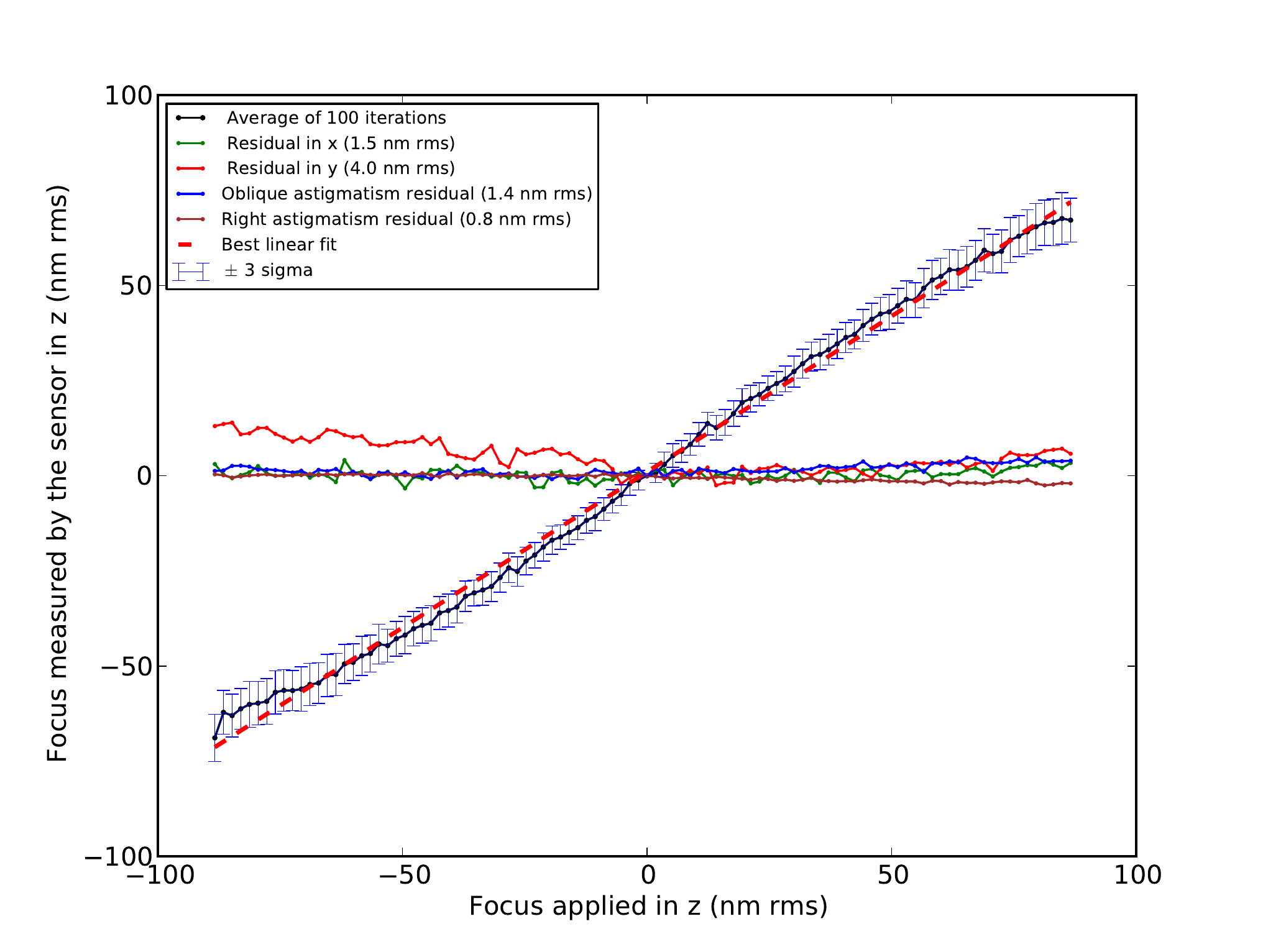}
   \end{tabular}
   \end{center}
   \caption[example] 
   { \label{fig:fig5}  Linear response of the sensor to the Focus aberration: X-axis plots the focus within $\pm$ 85 nm rms applied to the DM. Y-axis plots the focus measured by the sensor. The red dash line shows the best linear fit over the full measurement range with non-linearity percentage of 3.46\%. The residuals of the other low order aberrations average around zero indicating that there is negligible cross coupling. The experiment is repeated 100 times and the dispersion of the sensor measurements is thus shown as a $\pm~3 \sigma$ error bar. }
\end{figure}

\begin{figure}
   \begin{center}
   \begin{tabular}{c}
   \includegraphics[height=8.4cm]{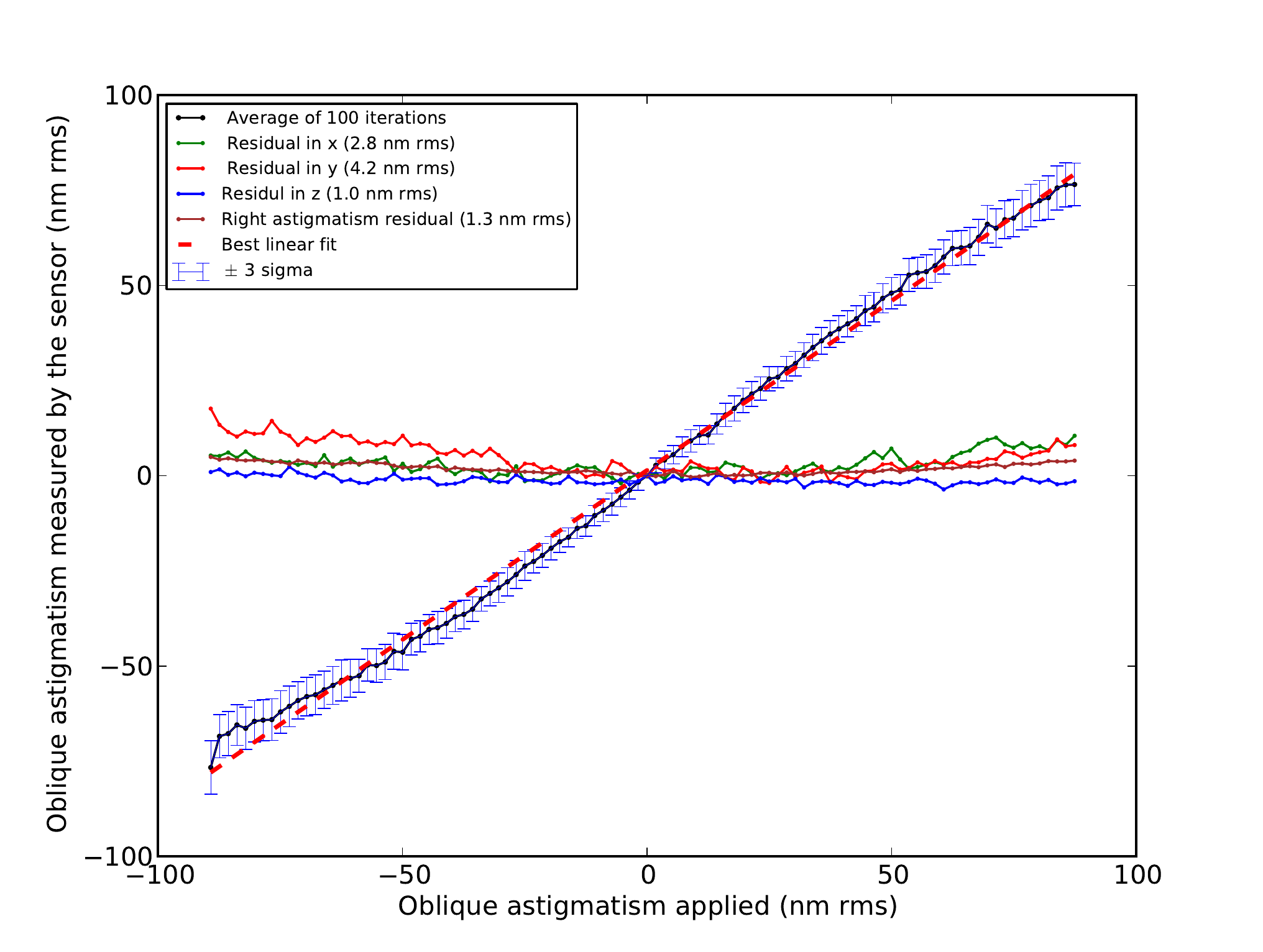}
   \end{tabular}
   \end{center}
   \caption[example] 
   { \label{fig:fig6} Linear response of the sensor to the Oblique astigmatism aberration: X-axis plots the oblique astigmatism within $\pm$ 85 nm rms applied to the DM. Y-axis plots the oblique astigmatism measured by the sensor. The red dash line shows the best linear fit over the full measurement range with non-linearity percentage of 3.1\%. The residuals of the other low order aberrations average around zero indicating that there is negligible cross coupling. The experiment is repeated 100 times and the dispersion of the sensor measurements is thus shown as a $\pm~3 \sigma$ error bar.}
\end{figure}

\begin{figure}
   \begin{center}
   \begin{tabular}{c}
   \includegraphics[height=8.4cm]{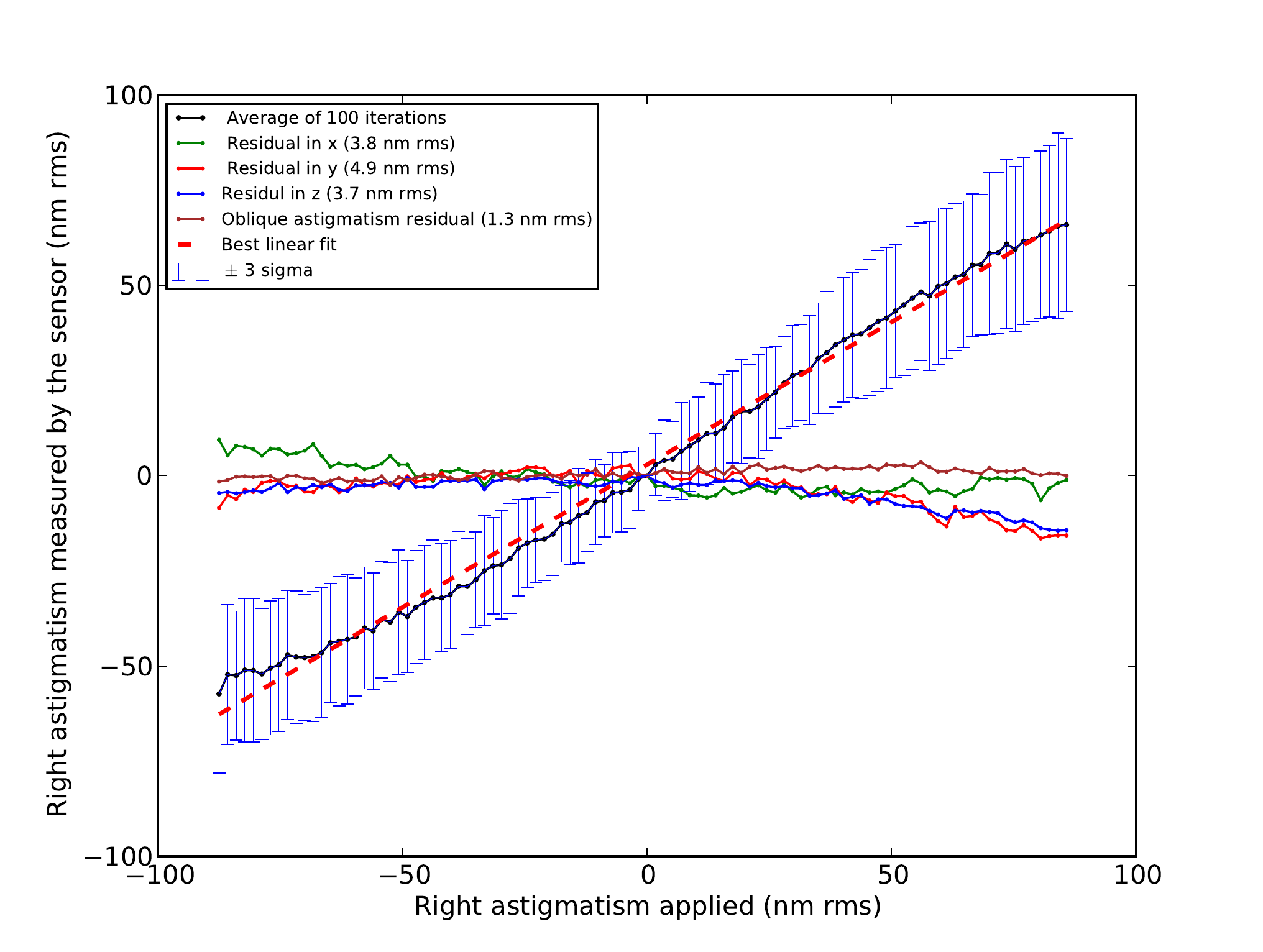}
   \end{tabular}
   \end{center}
   \caption[example] 
   { \label{fig:fig7}Linear response of the sensor to the Right astigmatism aberration: X-axis plots the right astigmatism within $\pm$ 85 nm rms applied to the DM. Y-axis plots the oblique astigmatism measured by the sensor. The red dash line shows the best linear fit over the full measurement range with non-linearity percentage of 3.58\%. The residuals of the other low order aberrations average around zero indicating that there is negligible cross coupling. The experiment is repeated 100 times and the dispersion of the sensor measurements is thus shown as a $\pm~3 \sigma$ error bar.}
\end{figure}

Figure \ref{fig:fig4}, \ref{fig:fig5}, \ref{fig:fig6} and \ref{fig:fig7} show the response of the sensor with respect to the aberrations applied for example: tip/tilt, focus, oblique and right astigmatisms respectively. Note that the experiment is repeated 100 times for each mode to reduce the low and high frequency noise. Each measurement is an average of 100 data points. The error bars in each graph show the noisy measurement that we have not quantified in this paper. The origin of the noise in the measurements can be DM non-linearity, photon noise, low order sensor read-out noise, drift in the reference image and local vibration of the optics. 

We show that LLOWFS can measure aberrations other than just tip-tilt. LLOWFS's response to each aberration is distinguishable from other modes and there is negligible cross coupling in the measurement when each aberration is studied independently. For example, in Figure \ref{fig:fig4}, the plot on the x-axis represents the tip aberration applied to the system and the y-axis represents the tip measured by the sensor. It is clear from the graph that apart from the response being linear to the tip mode, the residual in focus, oblique and right astigmatism are on average around zero, indicating little cross coupling.  

However we found out that the linearity range for each aberration is different. The sensor showed a non-linearity of 4.7\% over the range of $\pm$ 35~nm rms ($\pm$ 70 nm rms on the wavefront) for tip and tilt modes. Focus, oblique and right astigmatism showed a non-linearity of $<$ 5\% over the full measurement range of $\pm$ 85~nm rms ($\pm$ 170 nm rms on the wavefront ). We have not studied the origin of the noise sources yet, hence we can not explain the behavior of the sensor in detail in this paper.  

\subsection{Early on-sky Results}
\label{sec:onsky}

\begin{figure}[h]
   \begin{center}
   \begin{tabular}{c}
   \includegraphics[height=2.8cm]{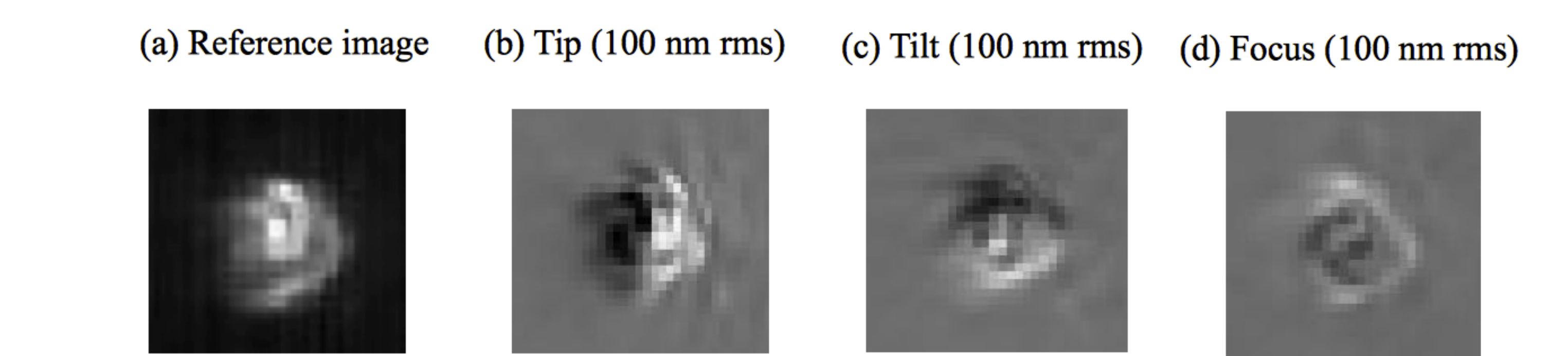}
   \end{tabular}
   \end{center}
   \caption[example] 
   { \label{fig:fig8}  On-sky images: (a) Reference image. Calibration frames for (b) Tip mode. $( c)$ Tilt mode. (d) Focus mode. All the calibration frames are obtained by applying 100 nm rms (0.24 $\lambda$/D angle on sky) phasemap to the DM for each mode individually. We used these calibration frame to close the loop on-sky. All the images are at same brightness scale.
}
\end{figure}

We have tested the LLOWFS initial closed loop on-sky performance with the PIAA and the VVC.  We report here the results with the VVC only on tip, tilt and focus. The AO188 closed the loop with a seeing of 1.3 arc seconds at H band providing strehl of $\approx$ 30 \%. The LLOWFS then acquired the reference frame on-sky as shown in the Figure \ref{fig:fig8} (a). 100 nm rms phasemap of tip, tilt (0.24 $\lambda$/D angle on sky) and focus is applied on the DM individually and the response matrix for each mode is obtained as shown in Figure \ref{fig:fig8}(b-d). 

\begin{figure}[h]
   \begin{center}
   \begin{tabular}{c}
   \includegraphics[height=7.2cm]{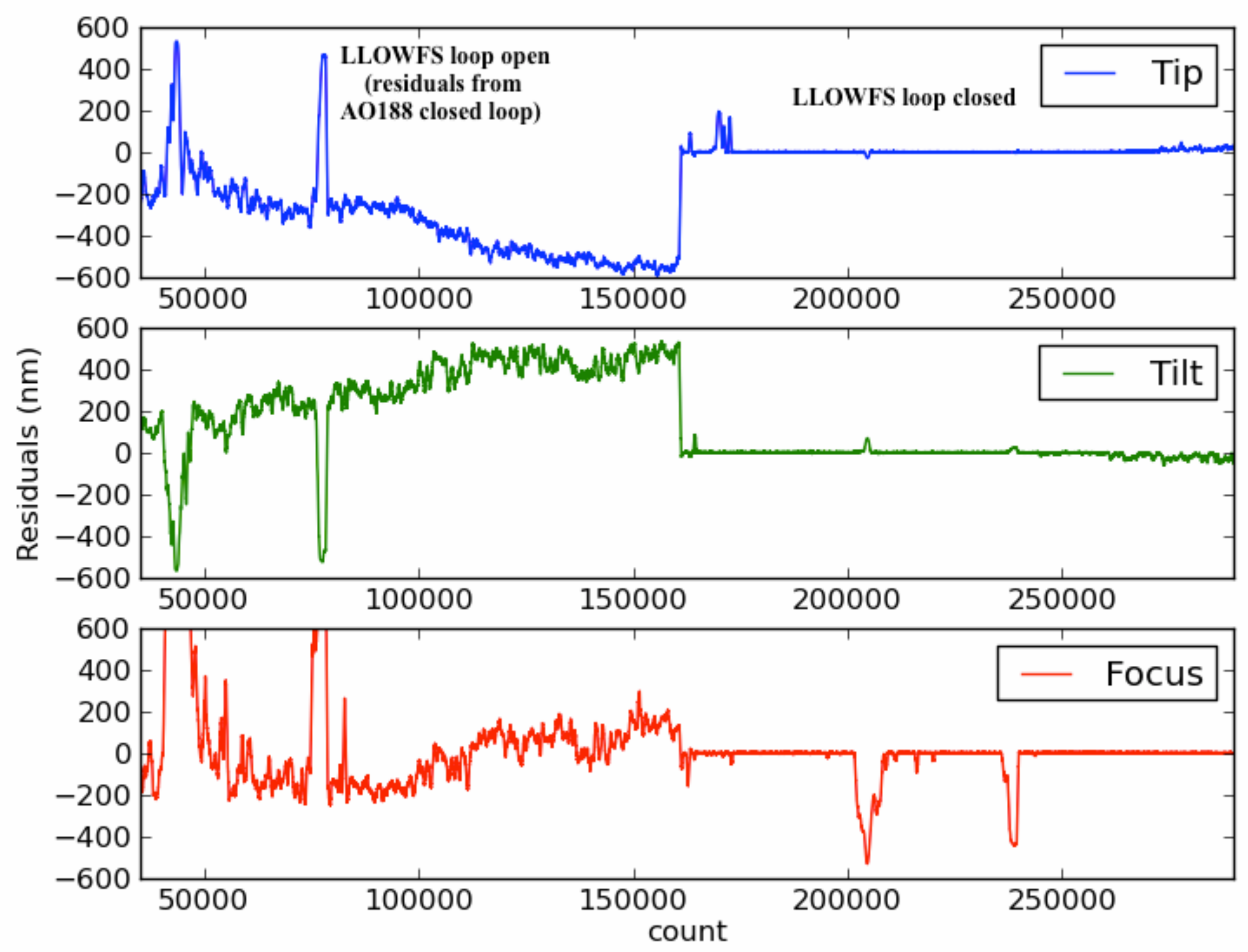}
   \end{tabular}
   \end{center}
   \caption[example] 
   { \label{fig:fig9}  Initial closed loop on-sky residuals obtained with LLOWFS on SCExAO at Subaru Telescope (April 2014). The coronagraph used is the VVC. The X axis shows the data binned by 100 frames. The Y axis shows the AO188 post wavefront residuals when the LLOWFS loop is open (count from 0 to 160000) and residuals for tip, tilt and focus when LLOWFS closed the low order loop (count from 160000 and onwards). The closed loop measurement accuracy per mode is $<$~7x10$^{-3}$~$\lambda$/D. The dips in the residuals of focus when the LLOWFS loop is closed is because of the instability in AO188 loop.}
\end{figure}

The LLOWFS sensed the on-sky aberrated wavefront reflected by the RLS and by using equation \ref{eq:1}, estimated the amount of the tip, tilt and focus present in the post AO188 corrected wavefront. The control matrix was computed by the singular value decomposition method and the commands were sent to the DM to compensate for the low order aberrations. The LLOWFS successfully closed the loop on the AO188 residuals at 80 Hz on Vega at 1.6~$\mu$m. The loop kept running for $\approx$ 1 hour with a gain of 0.3. The LLOWFs loop broke temporarily when AO188's loop became unstable because of the bad seeing. In Figure \ref{fig:fig9}, we obtained the closed loop residuals of 2.8 nm rms, 7.7 nm rms and 2.2 nm rms for tip, tilt and focus respectively at the time when LLOWFS loop was running. 
\section{Conclusion}
\label{sec:con}

We demonstrated the implementation and the initial performance of Lyot-based low order wavefront sensor (LLOWFS) on SCExAO in the laboratory as well as on-sky with the vector vortex coronagraph (VVC). The LLOWFS\cite{6} previously measured the tip-tilt only but in this paper we showed that LLOWFS is capable of measuring other low order modes such as focus, oblique and right astigmatisms. Moreover the measurement of the low order modes are independent from each other and there is a negligible cross correlation. LLOWFS initial closed loop on-sky performance on SCExAO at Subaru Telescope demonstrated that by extracting the residual starlight at the Lyot plane, it is possible to sense and correct the low order wavefront errors without introducing non-common path errors. We closed the loop on sky on tip, tilt and focus for vector vortex coronagraph at 1.6~$\mu$m with the closed loop accuracy of $<$~7x10$^{-3}$~$\lambda$/D. The LLOWFS is not parametrized yet.  The study of the noise sources such as read out noise, photon noise, DM non-linearity and local vibration will help to improve the performance of the sensor. SCExAO's Pyramid higher order wavefront corrections in visible will also enhance LLOWFS close loop performance in infrared by reducing the high frequency jitter. Future work includes the coupling of SCExAO's high order loop with the LLOWFS loop in order to achieve strehl ratio of $\approx$ 80 \% at H band. The coupling of the control loops will introduce the differential tip tilt errors in the infrared channel which we aim to correct with LLOWFS by changing the zero point of the Pyramid wave front sensor in visible. The working and performance of LLOWFS will be tested and verified with other phase mask coronagraphs on SCExAO such as four quadrant phase mask, eight octant phase mask and PIAACMC.

\end{document}